\documentclass[submitting]{nst}

\usepackage{subfigure,dcolumn}
\usepackage[T2A,T1]{fontenc}
\usepackage[russian,english]{babel}
\usepackage{float}
\usepackage{mathrsfs} 
\usepackage{amsmath}
\usepackage{url}

\usepackage{listings}
\makeatletter
\providecommand\orig@linenumbers{} 
\let\orig@linenumbers\linenumbers
\renewcommand\linenumbers{\relax}

\makeatother
\nolinenumbers

\begin{document}

\title{Lattice design of a storage-ring-based light source for  generating high-power fully coherent EUV radiation}\thanks{Supported by Shanghai Municipal science and Technology Major Project}
\author{Yu-Jie Lu}
\affiliation{Zhangjiang Laboratory, Shanghai 201210, China}
\author{Ao Liu}
\affiliation{School of Physical Science and Technology, ShanghaiTech University, Shanghai 201210, China}
\author{Chang-liang Li}
\affiliation{Shanghai Advanced Research Institute, Shanghai 201210, China}
\author{Kun Wang}
\affiliation{Shanghai Advanced Research Institute, Shanghai 201210, China}
\affiliation{Shanghai Institute of Applied Physics, Chinese Academy of Sciences, Shanghai 201800, China}
\affiliation{University of Chinese Academy of Sciences, Beijing 100049, China}
\author{Qing-lei Zhang}
\affiliation{Shanghai Advanced Research Institute, Shanghai 201210, China}
\author{Wei-shi Wan}
\affiliation{Quantum Science Center of Guangdong-Hong 
Kong-Macao Greater Bay Area, Shenzhen 518045, China}
\author{Wei-jie Fan}
\affiliation{Zhangjiang Laboratory, Shanghai 201210, China}
\author{Jun-hao Liu}
\affiliation{School of Physical Science and Technology, ShanghaiTech University, Shanghai 201210, China}
\author{Rui-chun Li}
\affiliation{Zhangjiang Laboratory, Shanghai 201210, China}
\author{Yan-xu Wang}
\affiliation{Shanghai Institute of Applied Physics, Chinese Academy of Sciences, Shanghai 201800, China}
\affiliation{University of Chinese Academy of Sciences, Beijing 100049, China}
\author{Kong-long Wu}
\affiliation{Zhangjiang Laboratory, Shanghai 201210, China}
\author{Ji Li}
\email[Corresponding author, ]{Ji Li, liji@zjlab.ac.cn}
\affiliation{Zhangjiang Laboratory, Shanghai 201210, China}
\author{Chao Feng}
\email[Corresponding author, ]{Chao Feng, fengc@sari.ac.cn}
\affiliation{Shanghai Advanced Research Institute, Shanghai 201210, China}

\begin{abstract}
We present the physical design and systematic optimization of a high-performance storage ring tailored for the generation of high-power coherent radiation, with particular emphasis on the extreme ultraviolet (EUV) regime. The proposed ring adopts a Double Bend Achromat (DBA) lattice configuration and integrates 12 superconducting wigglers to significantly enhance radiation damping and minimize the natural emittance. And a bypass line is adopted to generate high power coherent radiation. Comprehensive linear and nonlinear beam dynamics analyses have been conducted to ensure beam stability and robustness across the operational parameter space. The optimized design achieves a natural emittance of approximately 0.8 nm and a longitudinal damping time of around 1.4 ms, enabling the efficient buildup of coherent radiation. Three-dimensional numerical simulations, incorporating the previously proposed angular dispersion-induced microbunching (ADM) mechanism, further confirm the system’s capability to generate high-power EUV coherent radiation, with output powers reaching the order of several hundred watts. These results underscore the strong potential of the proposed design for applications in coherent photon science and EUV lithography.
\end{abstract}


\maketitle

\section{Introduction}

\begin{figure*}[!htb]
\includegraphics
  [width=0.9\hsize]
  {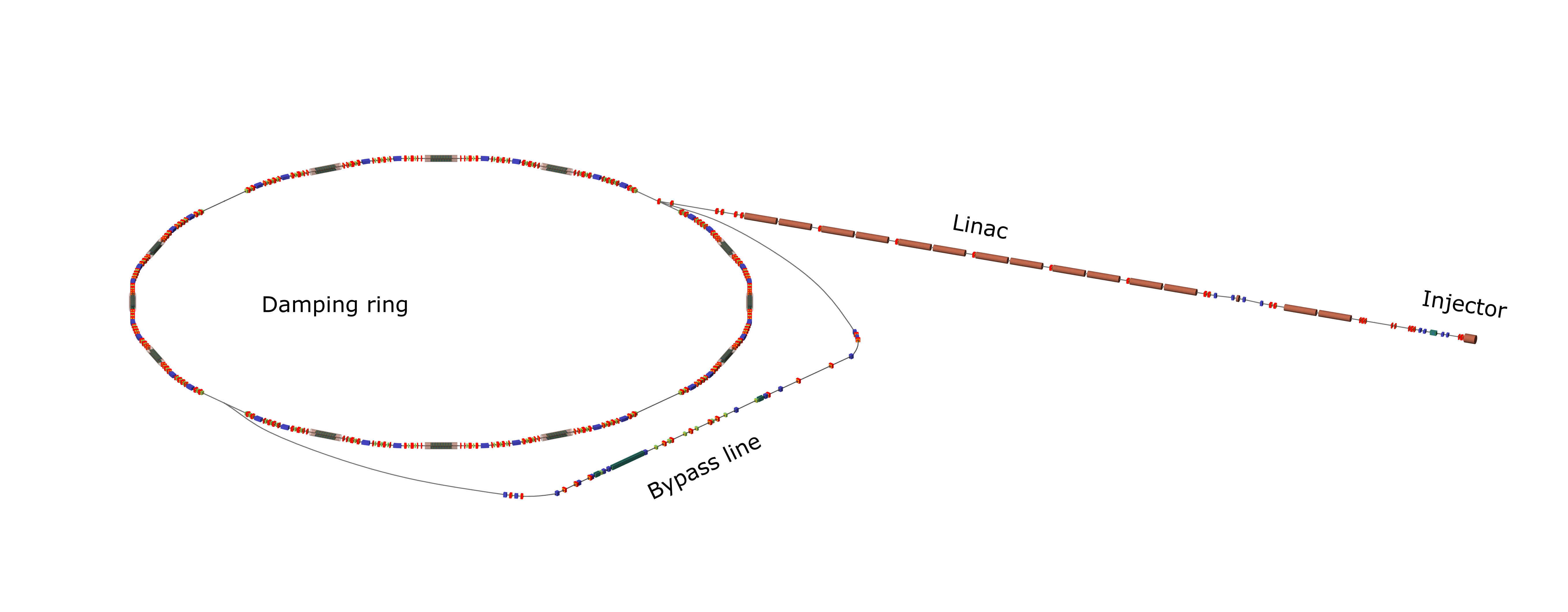}
\caption{Schematic layout of storage ring light source based on the angular dispersion-induced microbunching.}
\label{fig:1}
\end{figure*}

Accelerator-based light sources can be broadly categorized into synchrotron radiation sources based on storage rings and free-electron lasers (FELs) driven by linear accelerators. FELs provide ultrashort pulses, high peak power, and access to short wavelengths, making them indispensable for time-resolved and high-field experiments. However, their repetition rate is typically limited to the range of 100 Hz to 100 kHz due to constraints from the electron gun and RF system, thereby limiting average brightness and photon flux. Furthermore, FELs often require strong bunch compression to achieve high peak currents, resulting in broad spectral bandwidths that are not well suited for experiments demanding high energy resolution, such as high resolution ($\sim$0.1 meV) angle-resolved photoemission spectroscopy (ARPES)~\cite{Sobota2021}.

In contrast, storage rings inherently operate at high repetition rates with relatively long and stable electron bunches. Combining the advantages of FELs and storage rings offers a promising path to a high repetition rate, coherent light source with enhanced average power and energy resolution. Although Self-Amplified Spontaneous Emission (SASE)-based storage ring FELs ~\cite{Billardon1983,kim1985issues,litvinenko1996duke,wu1997performance,Cai2013,Di2018,shen2021,Cai2025} have been investigated, their performance is fundamentally constrained by the relatively large energy spread of the stored beam. Achieving sufficient radiation power under these conditions requires long undulator sections, which significantly increase the overall size and construction cost of the facility. To mitigate these limitations, alternative schemes originally developed for FELs—such as coherent harmonic generation (CHG) ~\cite{PhysRevLett.53.2405} and echo-enabled harmonic generation (EEHG) ~\cite{stupakov2009using,PhysRevSTAB.12.030702,Evain_2012,Liu2018,LiuW2019,Hwang2020}, both relying on laser-based electron beam manipulation—have been proposed for adaptation to storage-ring-based light sources. However, these methods typically introduce large energy modulations to the beam, resulting in degraded beam quality and a substantial reduction in usable repetition rate—thereby compromising one of the fundamental advantages of storage rings.

To overcome these limitations and enable high-average-power, narrow-band, fully coherent EUV radiation, several alternative approaches have been proposed, including steady-state microbunching (SSMB) ~\cite{Ratner2010,Tang2018,deng2021experimental,zhang2021ultralow,Tang2022ssmb,Li2023,Deng2024} proposed by Tsinghua University team and angular dispersion-induced microbunching (ADM) based storage ring light source~\cite{Feng2017,Changliang2020,yujie2022,Jiang2022} proposed by Shanghai Synchrotron Radiation Facility. In the SSMB scheme, the laser modulator forms potential buckets on the scale of the laser wavelength, serving a role analogous to that of the radio-frequency (RF) cavity in conventional storage rings. Since the laser wavelength is much shorter than that of the RF cavity, the laser-induced potential well forms ultrashort electron bunches with high charge density. The precision-tailored lattice ensures that the microbunches are stored in a steady state within the storage ring. Once realized, this approach has the potential to generate high-average-power, fully coherent radiation from the EUV to the x-ray regime.

Another alternative approach is the ADM scheme for storage ring light sources, which leverages the intrinsically small vertical emittance of the electron beam. Ref.~\cite{Jiang2022} proposes a scheme utilizing a storage ring equipped with a bypass line. This configuration is the focus of this paper. Within the bypass, microbunches formed via transverse-longitudinal coupling generate high-harmonic coherent radiation before completely dissipating. Thus, the main storage ring remains free of microbunches and functions essentially as a damping ring.

The layout of a fully coherent light source based on the ADM concept is illustrated in Fig.~\ref{fig:1}. An electron beam with ultra-low vertical emittance and high peak current from the storage ring is extracted into the bypass line, where it forms microbunches and subsequently emits coherent EUV radiation. The vertical emittance grows in the bypass; consequently, the beam is reinjected into the storage ring. There, its quality is rapidly restored via strong synchrotron radiation damping until it again meets the requirements of ADM, after which it can be extracted to the bypass, thereby enabling a continuous process. A decrease in bunch charge is compensated by the full-energy linac , ensuring  stable output power of the coherent EUV radiation.

Building upon the work in Ref.~\cite{Jiang2022}, we propose a new, practical, and robust lattice design for the damping ring and bypass line. The remainder of this paper is structured as follows. Section II details the lattice design of the damping ring, addressing both linear and non-linear beam dynamics. Section III is dedicated to the lattice design of the bypass line. The integration of these two components and an evaluation of the overall radiation performance are presented thereafter. Finally, Section IV provides a concise summary.

\section{Damping ring design}\label{sec:artwork}

The project of a fully coherent storage-ring–based light source requires a storage ring with high beam current, low emittance, and strong radiation damping to deliver high-quality electron beams for the bypass line, thereby enabling the generation of high-average-power EUV radiation. In the preliminary design phase, a candidate lattice design was proposed in Ref.~\cite{Jiang2022}. The initial configuration adopts an eight-period triple-bend achromat (TBA) scheme, in which four TBA cells form a superperiod. The straight sections on both sides of each superperiod are designed with high beta functions, while the inner straight sections adopt a low-beta design. Each low-beta straight section accommodates an insertion device composed of three superconducting sandwich-type damping wigglers to facilitate the
I transformation, which serve to enhance the synchrotron radiation losses and thereby significantly reduce the damping times. In total, the ring hosts 18 superconducting damping wigglers distributed across the lattice.

Toward the technical design stage, we present in this section a more practical and robust damping ring lattice, aiming to address the aforementioned challenges while preserving the performance goals for coherent radiation generation. In the following subsections, the linear optics design and nonlinear dynamics are discussed in detail.

\subsection{Linear optics design}
According to the definition of the damping time:
\begin{equation}\label{1}
\tau_y= 2\frac{T_0E_0}{U_0},
\end{equation}
where $U_0 = \frac{C_r}{2\pi}E_0^4I_2$ with $C_\gamma = \frac{4\pi}{3}\frac{r_e}{(m_ec^2)^3} = 8.85\times 10^{-5} \frac{m}{GeV^2}$, c the speed of light in free space, $r_e$ the classical electron radius, $m_e$ the electron mass, $\rho$ is the bending radius, $E_0$ is the beam energy and $I_2 = \oint \frac{1}{\rho^2} \, ds$. The damping time is primarily determined by the synchrotron radiation losses. In high-field damping wigglers, the bending radius $\rho$ of the electron trajectory is much smaller than that in the dipole magnets, leading to the radiation integral $I_2$ from the wigglers that is significantly larger than the contribution from the dipoles. Consequently, increasing the fraction of damping wigglers in the lattice can effectively reduce the damping time. In addition, the equilibrium emittance can be expressed as: 
\begin{equation}
    \epsilon_0 = C_q \frac{\gamma^2 I_5}{j_x I_2},
\end{equation}
where
\begin{equation}\label{Hx}
\begin{aligned}
    \quad I_5 &= \oint \frac{H_x}{|\rho|^3} \, ds, \\
    j_x&\approx 1,\\
    \quad H_x &= \gamma_x \eta_x^2 + 2\alpha_x \eta_x \eta_{px} + \beta_x\eta_{px}^2,
\end{aligned}
\end{equation}
where $C_{q} = \frac{55}{48 \sqrt{3}} \, \frac{r_{e} \hbar}{m_{e}}$ with $\hbar$ the reduced Planck’s constant, $\gamma$ is the Lorentz factor. The equilibrium emittance is primarily determined by the ratio $I_5/I_2$. In high-field damping wigglers, the radiation integral $I_2$ is much larger than that in the dipole magnets. Since the radiation integral $I_5$ contains the dispersion-dependent function H, its contribution remains small (almost unchanged) when the wigglers are located in dispersion-free regions. Therefore, by installing a sufficient number of damping wigglers in the low-dispersion sections of the storage ring, the emittance can also be effectively reduced.

In summary, since one of the key design objectives is to achieve a vertical damping time below 1.5 ms, a large number of superconducting damping wigglers are required to substantially enhance the contribution of $I_2$. As these wigglers must be placed in dispersion-free straight sections with low beta functions, their contribution to $I_5$ can be neglected. Therefore, both the damping time and the equilibrium emittance of the damping ring are essentially dominated by the superconducting wigglers. Based on this conclusion, our core design strategy is to maximize the fraction of $I_2$ contributed by the wigglers in the storage ring. Accordingly, the bare lattice only needs to emphasize compactness and robustness, without being strictly constrained by emittance-oriented lattice structures. Considering these requirements, a natural choice is to adopt the simple double-bend achromat (DBA) lattice combined with superconducting wigglers, which maximize the contribution of the wigglers to strong damping effect while offering a simple lattice structure and favorable nonlinear beam dynamics.

\begin{figure}[H]
\centering
\includegraphics[width=1\hsize]{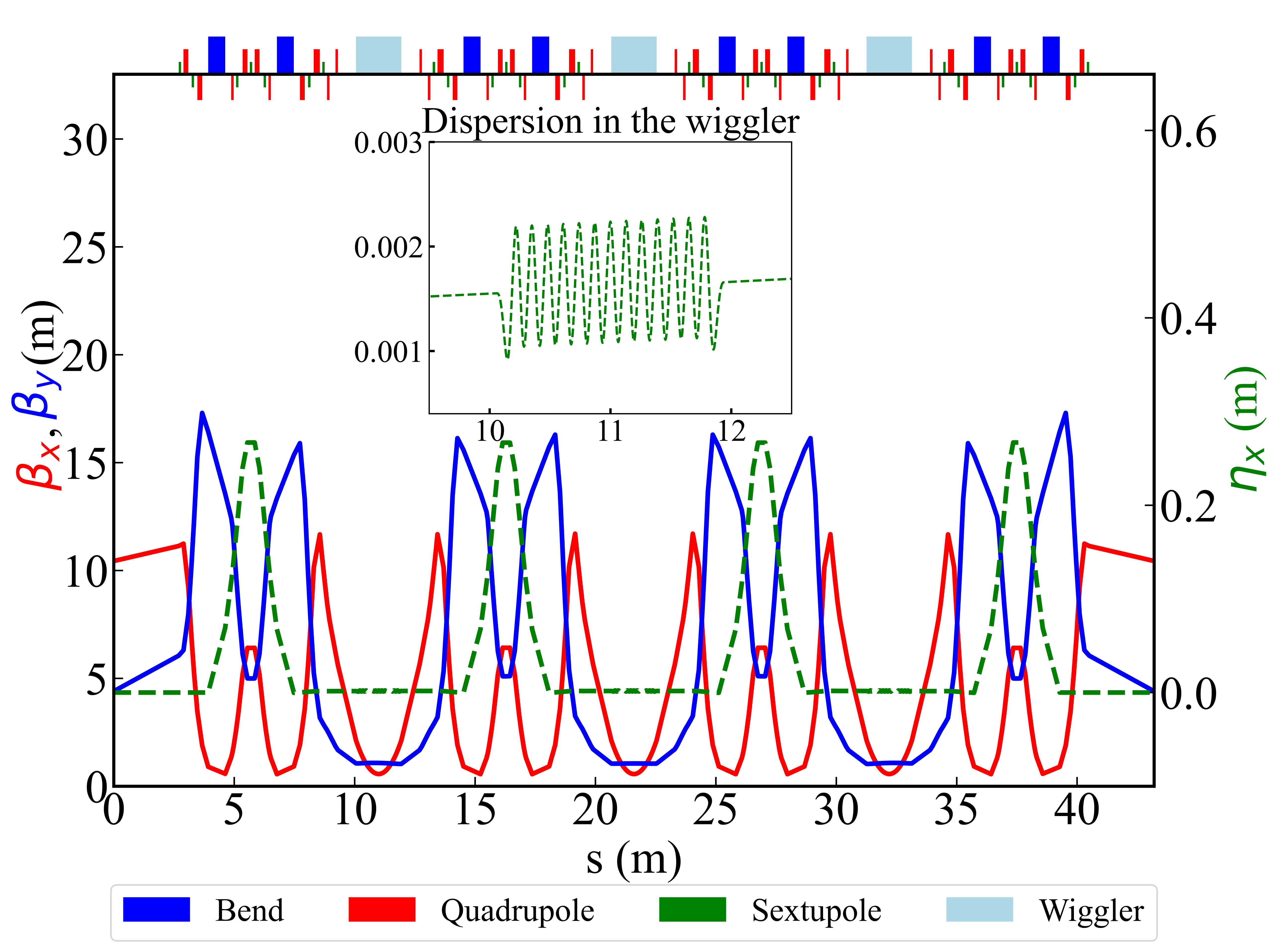}
\caption{Beta and dispersion functions for one quarter of the ring.}
\label{fig:22}
\end{figure}

Furthermore, the storage ring must maintain a compact layout with the consideration of damping performance and total cost, while simultaneously offering adequate straight sections for beam injection, extraction, and RF cavities. In this design, we therefore adopt four straight sections. Based on this consideration, several lattice configurations were studied, including DBA, DBA-wiggler-DBA, DBA-wiggler-DBA-wiggler-DBA and DBA-wiggler-DBA-wiggler-DBA-wiggler-DBA, where the wigglers are located in the straight sections between DBA arcs. Using the beam parameters of the preliminary design listed in Ref.~\cite{Jiang2022}, evaluations of both the damping times and the equilibrium emittance were carried out. Furthermore, taking into account the practical challenges in wiggler design and realization, each superconducting wiggler was assumed to have a length of 1.8 m and a peak magnetic field of 6.5 T, which is a reasonable assumption. As a starting point, the theoretical minimum emittance of a single DBA structure~\cite{Teng1985,Wolski2014} can be expressed as:
\begin{equation}\label{Hx}
\begin{aligned}
\epsilon_{0,DBA,min}=C_q\gamma^2\frac{I_{5,DBA}}{j_xI_{2,DBA}}\approx \frac{1}{4\sqrt{15}}C_q\gamma^2\theta^3.
\end{aligned}
\end{equation}
To maximize the damping effect, the magnetic field strength of dipole magnets were chosen to be 1.4 T which operate at the highest feasible magnetic field strength, corresponding to a bending radius of $\rho_{DBA}=2.38$ m. For the contribution of the damping wigglers, the emittance can be expressed as:
\begin{equation}\label{Hx}
\begin{aligned}
\varepsilon_{0,DBA}=C_q\gamma^2\frac{I_{5w}}{j_xI_{2w}}.
\end{aligned}
\end{equation}
where the synchrotron radiation integrals for the wigglers, $I_{2w}$ and $I_{5w}$, can be simplified as:
\begin{equation}\label{Hx}
\begin{aligned}
I_{2w} = \int_{0}^{L_w} \frac{1}{\rho^2} \, ds 
= \frac{1}{(B\rho)^2} \int_{0}^{L_w} B^2 \, ds
= \frac{1}{(B\rho)^2} \cdot \frac{B_w^2 L_w}{2}
\end{aligned}
\end{equation}
\begin{equation}\label{Hx}
\begin{aligned}
I_{5w} \approx \frac{4}{15\pi}\frac{\langle\beta_x\rangle L_w}{\rho_w^5k_w^2}.
\end{aligned}
\end{equation}
where $L_w$ denotes the total length of the wiggler, $B\rho$ is the magnetic rigidity, $B_w$ is the peak magnetic field of the wiggler, and $\rho_w$ is the bending radius of the electron trajectory inside the wiggler. The wiggler wave number is given by $k_w=2\pi/\lambda_w$, with $\lambda_w$ being the wiggler period length. The quantity $\langle\beta_x\rangle$ represents the average horizontal beta function within the wiggler, and in the present calculation we set $\langle\beta_x\rangle=1$ m. In our survey, the wiggler period was chosen to be $\lambda_w=130$ mm. The wiggler synchrotron-radiation integrals determined for this choice are $I_{2w}=3.42$ and $I_{5w}=1.844×10^{-3}$. Consequently, the horizontal equilibrium emittance of the DBA lattice combined with wigglers can be written as:
\begin{equation}\label{8}
\varepsilon_x = \frac{C_q \gamma^2}{j_x}\frac{I_{5,\mathrm{DBA}} + I_{5w}}{I_{2,\mathrm{DBA}} + I_{2w}},
\end{equation}
According to the Eq.~\ref{1}, the vertical damping time of the wiggler can be written as:
\begin{equation}\label{9}
\tau_y= 2\frac{4\pi T_0}{C_\gamma E_0^3(I_{2w}+I_{2,DBA})},
\end{equation}
Using Eqs.~\ref{8} and~\ref{9}, we perform a estimate of the equilibrium emittance and the vertical damping time for several DBA–wiggler configurations. The results are summarized in Table~\ref{tab:DBAwig}.
\begin{table*}[!htb]
\centering
\caption{Estimated beam parameters for different DBA-wiggler configurations.}
\label{tab:DBAwig}
\begin{tabular*}{10cm} {@{\extracolsep{\fill} } llr}
\toprule
  & $\tau_y$(ms) & Emittance(nmrad) \\
\midrule
DBA  & 26.7 & 45.87 \\
DBA-wig-DBA  &  4.3  &  1.59 \\
DBA-wig-DBA-wig-DBA  & 2.35 & 0.87 \\
DBA-wig-DBA-wig-DBA-wig-DBA & 1.61  & 0.786 \\
\bottomrule
\end{tabular*}
\end{table*}
From the table, it can be seen that as the fraction of the wiggler within a single superperiod increases, the damping time can be significantly reduced, and the emittance can also decrease. However, as the wiggler fraction increases, the rate of emittance reduction gradually slows down, while the associated costs continue to rise. Moreover, an excessive number of DBA cells not only enlarges the ring circumference but may also leads to small momentum compaction factor, thereby affecting the momentum acceptance. We therefore selected the last combination. Based on this configuration, we employed a thin-dipole model (Appendix~\ref{app:dipole}) to represent the wiggler and performed an overall optimization of its length, number of periods, and peak field. The final design specifies a total wiggler length of 1.885 m, 14.5 periods, a period length of 130 mm, and a peak magnetic field of 6.3135 T. To ensure an undisturbed beam trajectory through the wiggler and a zero second field integral, the peak fields at the end poles are adjusted to 1.4837 T and 4.6404 T . Unlike ideal wiggler models, which are inherently achromatic, real wigglers have fields that start from zero, making them non-achromatic by nature. Therefore, the adjustment of the end-pole fields must also ensure that the wiggler as a whole remains achromatic. According to Eq. (8), the damping time is inversely proportional to the cube of the beam energy, $E_0^3$; therefore, increasing the beam energy can reduce the damping time and enhance the damping effect. In addition, higher beam energy mitigates collective effects such as intrabeam scattering (IBS). Through optimization, the beam energy was chosen to be 1.4 GeV. The final design adopts a superperiod structure composed of four DBA cells and three superconducting wigglers, with four superperiods around the entire ring. We then optimized the ring parameters with respect to emittance, damping time, circumference, working point, and momentum compaction factor. Fig.~\ref{fig:22} shows the $\beta$-functions and dispersion functions for one quarter of the storage ring (corresponding to a single superperiod), clearly illustrating the oscillations of the dispersion within the wiggler sections. The main ring parameters are summarized in Table~\ref{tab:1}. 

It is worth noting that, since this lattice extensively employs superconducting damping wigglers, the wiggler serves as a critical component of the lattice rather than merely an insertion device. Therefore, during the lattice design process, it is essential to fully account for the vertical focusing effect introduced by the wiggler and the impact of its induced dispersion on the emittance. We first need to compute the electron trajectory inside the wiggler based on its three-dimensional field distribution. A series of thin dipole magnets are then placed along this trajectory to ensure that they correctly reproduce the wiggler's vertical focusing effect, and to guarantee that the transfer matrix derived from this thin-dipole model matches the one obtained numerically from the 3D field distribution. This approach allows us to design the lattice based on the Courant-Snyder formalism~\cite{Courant1958} using a thin-dipole representation of the wiggler—a versatile and efficient method. Given that the wiggler is an s-dependent 3D magnetic element, the thin-dipole model provides a sufficiently accurate approximation, while the envelope method~\cite{Wolski2006,Wolski2014} offers a more rigorous alternative. Thus, upon completion of the design, the envelope method should be applied to verify its reliability.

\begin{table}[!htb]
\caption{Ring parameters adopted in simulations.}
\label{tab:1}
\begin{tabular*}{8cm} {@{\extracolsep{\fill} } llr}
\toprule
Beam energy (MeV)  & 1400 \\
Circumference (m)  &  172.8  \\
Tune (x/y)  & 19.14/9.23 \\
Horizontal emittance (nm·rad)  & 0.795\\
Energy spread & 0.123\%  \\
Energy loss per turn (MeV) &  1.1 \\
Damping time (x/y/s) (ms)  & 1.46/1.46/0.73\\
RF frequency (MHz) & 499.66\\
RF voltage (MV) & 3.8\\
Harmonic number &  288\\
Bunch charge (nC) &  2.88   \\
number of bunches & 200   \\
Bunch length (mm) & 2.89\\
Beam current (A) & 1\\
Peak current (A) & 120\\
Coupling ratio & 0.3\% \\
Touschek lifetime (h) & 1.11 \\
\bottomrule
\end{tabular*}
\end{table}

Compared with the Courant–Snyder theory, the beam envelope method obtains the required physical quantities by solving the eigenvalues and eigenvectors. In the present scheme, for the wiggler with a complex three-dimensional magnetic field, numerical tracking can be employed instead of relying on transfer matrices. Moreover, this method is capable of handling complicated coupled optics problems. First, the beam envelope method are applied to determine the critical physical parameters of the storage-ring light source equipped with high-field superconducting wigglers. The detailed derivation is presented in the Appendix ~\ref{app:envelope}.

\begin{figure*}[t]
\centering
\includegraphics[width=\textwidth]{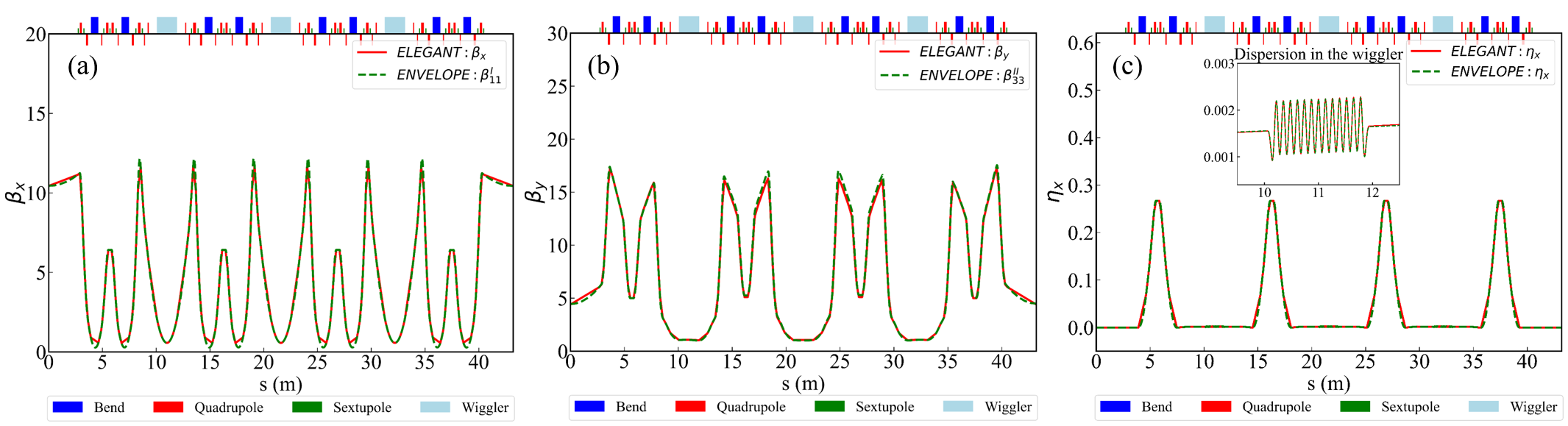}
\caption{Comparison of the Twiss functions, including the horizontal beta function $\beta_x$ (a), the vertical beta function $\beta_y$ (b), and the horizontal dispersion function $\eta_x$ (c), between the ELEGANT and envelope method incorporating three superconducting wigglers.}
\label{fig:2}
\end{figure*}

As a representative case, fully coherent storage ring light source design adopts a superperiod consisting of four DBA cells and three superconducting wigglers. The evolution of the generalized twiss matrices along the storage ring can be achieved in Eq.~\ref{eq:betaa}. In the uncoupled case, generalized twiss matrices can be reduced to the familiar Courant-Snyder form, as shown in Eqs.~\ref{eq:beta1},~\ref{eq:beta1},~\ref{eq:eta}. Fig.~\ref{fig:2} compares the Twiss functions of a single superperiod obtained from Courant-Snyder theory and from the beam envelope method. In the Courant–Snyder theory, the wiggler is represented using a thin-dipole model implemented in ELEGANT code~\cite{Borland2000}, while the beam envelope method is realized using a self-developed code. In addition to correctly reproducing the characteristic oscillations of the dispersion within the wiggler, the remaining Twiss functions exhibit excellent agreement between the two approaches.

In the next step, we will employ the envelope method in conjunction with the wiggler field map to compute the equilibrium emittance of the storage-ring light source. Detailed derivations are also presented in Appendix~\ref{app:envelope}.       

The equilibrium emittance can be achieved by solving Eq.~\ref{eq:B19}. In the storage ring, the eigen-emittance in the horizontal direction can be calculated from the solution for the sigma matrix (Eq. \ref{eq:sigma}) as: 0.753 nm$\cdot$rad. As shown in Table~\ref{tab:1}, the result obtained from the two methods are in very good agreement. This consistency validates the reliability of our numerical modeling.

\subsection{Nonlinear beam dynamics}
Despite the inherent nonlinear dynamic robustness of the DBA lattice, the extensive use of superconducting damping wigglers necessitates a detailed optimization of the nonlinear beam dynamics.

\begin{table}[!htb]
\caption{Ring parameters with IBS.}
\label{tab:IBS}
\begin{tabular*}{8cm} {@{\extracolsep{\fill} } llr}
\toprule
Beam energy (MeV)  & 1400 \\
Circumference (m)  &  172.8  \\
Tune (x/y)  & 19.14/9.23 \\
Horizontal emittance (nm·rad)  & 1.1\\
Energy spread & 0.129\%  \\
Energy loss per turn (MeV) &  1.1 \\
Damping time (x/y/s) (ms)  & 1.46/1.46/0.73\\
RF frequency (MHz) & 499.66\\
RF voltage (MV) & 3.8\\
Harmonic number &  288\\
Bunch charge (nC) &  2.88   \\
Bunch length (mm) & 2.89\\
Betatron coupling & 0.3\% \\
Touschek lifetime (h) & 1.32 \\
\bottomrule
\end{tabular*}
\end{table}

\begin{figure*}[!htb]
\includegraphics
  [width=0.9\hsize]
  {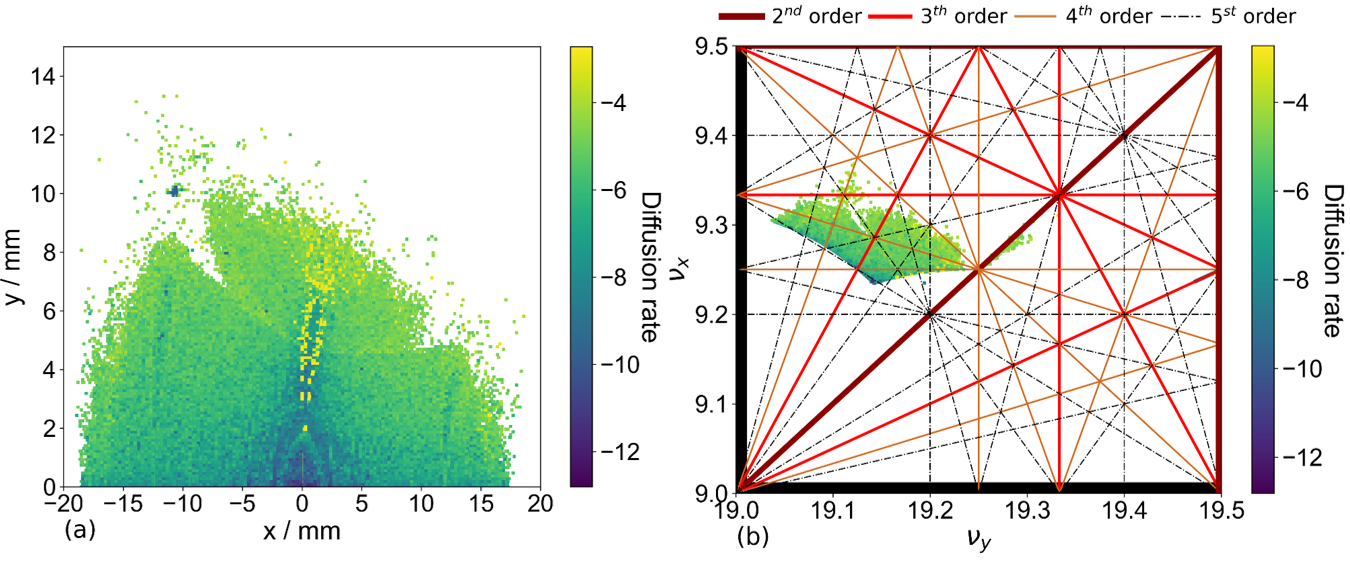}
\caption{ Frequency map analysis extracted from 1024-turn tracking using Elegant: (a) dynamic aperture with diffusion rates, (b) tune footprint.}
\label{fig:3}
\end{figure*}
In the optimization, the wiggler was modeled using the full 3D magnetic field distribution rather than the thin-dipole approximation.

The nonlinear dynamics optimization was carried out with Elegant. Throughout the optimization process, the chromaticities were consistently corrected at +1.5 in both transverse planes. Both sextupole strengths and their longitudinal positions were further optimized to maximize the dynamic aperture (DA), while constraints were imposed to preserve adequate spacing between magnets for engineering feasibility. The effective DA—the primary optimization objective—is defined as the phase-space area occupied by particles that survive 1024 turns of tracking in Elegant, with those crossing integer or half-integer resonances regarded as lost.

As shown in the frequency map of Fig.~\ref{fig:3}(a), the optimized sextupole configuration produces a large on-momentum DA of about $\pm 17$ mm. This result guarantees high injection efficiency for the off-axis injection scheme and demonstrates that the amplitude-dependent tune shifts (ADTS) are effectively suppressed. The tune footprint presented in Fig.~\ref{fig:3}(b) further confirms the compact tune shifts of the surviving particles.
\begin{figure}[H]
\centering
\includegraphics[width=1\hsize]{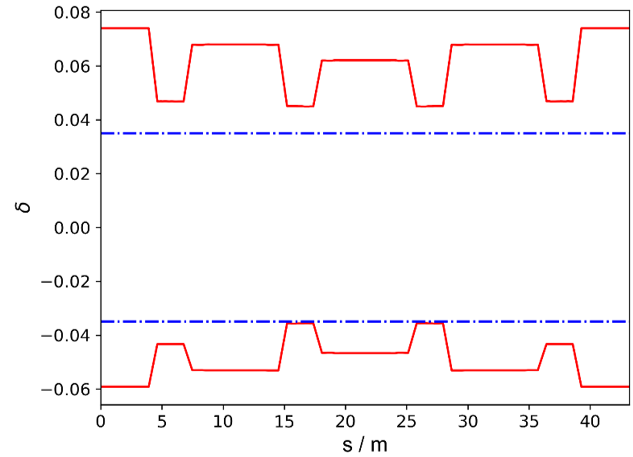}
\caption{Local momentum acceptance of one superperiod.}
\label{fig:4}
\end{figure}
Low vertical emittance and high peak current in the electron beam are essential for generating high-power EUV coherent radiation in a storage ring. This inevitably leads to a shortened Touschek lifetime. For low-energy, high-current storage rings, the local momentum aperture (LMA) is the limiting factor for the Touschek lifetime. Consequently, we performed a detailed optimization of the LMA, with particular emphasis on the DBA arc sections where the momentum aperture is most constrained by large disperison. Throughout this process, special attention was given to maintaining a sufficiently large DA. The LMA of one superperiod in the ring, obtained from tracking simulations with Elegant, is shown in Fig.~\ref{fig:4}. In the arc section, the LMA exceeds 3$\%$, which represents a relatively large value and is of particular importance for ensuring a sufficient Touschek lifetime.
\begin{figure}[H]
\centering
\includegraphics[width=0.8\hsize]{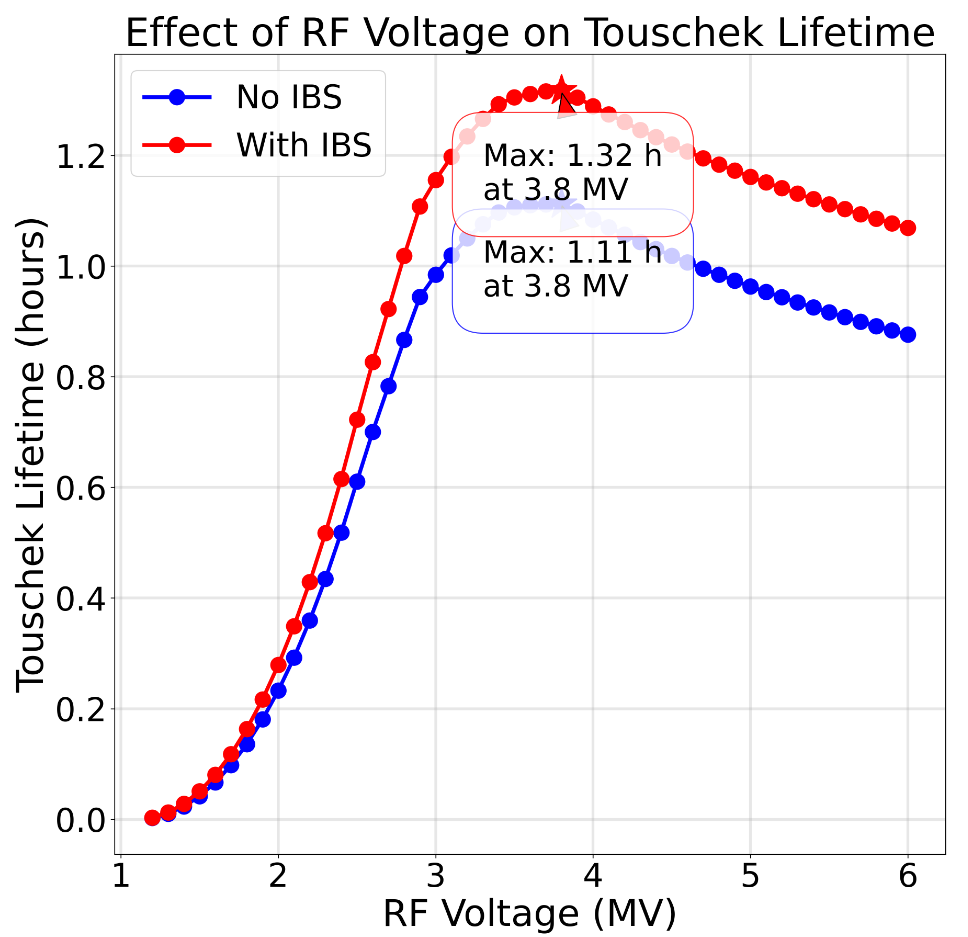}
\caption{Touschek lifetime vs RF voltage.}
\label{fig:ibs}
\end{figure}
Fig.~\ref{fig:ibs} shows the effect of RF voltage on the Touschek lifetime. As illustrated in the figure, the storage ring achieves its maximum Touschek lifetime when the RF voltage is set to 3.8 MV. Additionally, the intra-beam scattering (IBS) effect cannot be ignored due to the low emittance and high beam current. The beam parameters with considering IBS effects are shown in Table~\ref{tab:IBS}. The IBS effect is evaluated using ELEGANT based on Bjorken–Mtingwa’s formula~\cite{Borland2000}.

\section{Bypass line design}
\begin{figure*}[!htb]
\includegraphics
  [width=0.9\hsize]
  {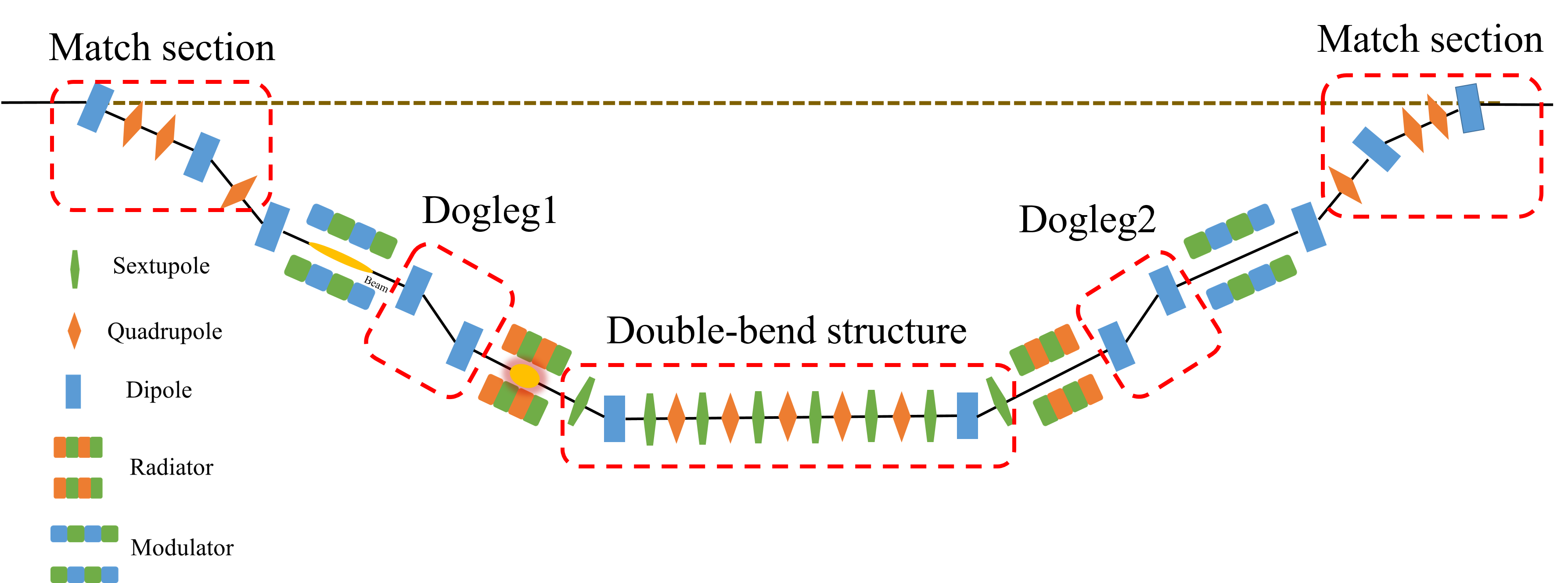}
\caption{Schematic layout of bypass line.}
\label{fig:bypass}
\end{figure*}
To generate the high-average-power EUV radiation, a bypass line is employed. In this section, a bypass line design will be introduced including modulation and demodulation techniques~\cite{Xiang2011,Ratner2011,Changliang2019,Weihang2024} based on ADM scheme. Fig.~\ref{fig:bypass} shows the schematic layout of the bypass line. The basic layout is derived from the preliminary design~\cite{Jiang2022}. According to the optimized conditions of the ADM scheme, an extremely small vertical angular dispersion of the electron beam at the entrance of the ADM section is required in order to achieve a large bunching factor and thus generate high-peak-power radiation pulse. To meet this requirement, the beam first passes through a matching section to provide a large beta function at the entrance of ADM. It then traverses a dipole magnet which generates dispersion, followed by a modulation undulator where energy modulation is introduced. After passing through a dogleg, the beam forms microbunching, and coherent radiation is produced in the radiator. Since the energy modulation process leads to an increase in energy spread and introduces vertical dispersion, vertical emittance growth can occur through transverse–longitudinal coupling. To achieve a high repetition rate and generate high-average-power radiation, the electron beam must be injected into the ring for damping, so that it can reach the beam quality required for bypass line. Consequently, a demodulation section is introduced following the ADM scheme to restore the beam properties as quickly as possible, thus enabling the generation of high-average-power radiation. In the following subsection, we will present the optical design of the entire bypass line as well as an evaluation of the radiation performance based on this design.
\subsection{Lattice design of bypass line}
\begin{figure}[H]
\centering
\includegraphics[width=0.8\hsize]{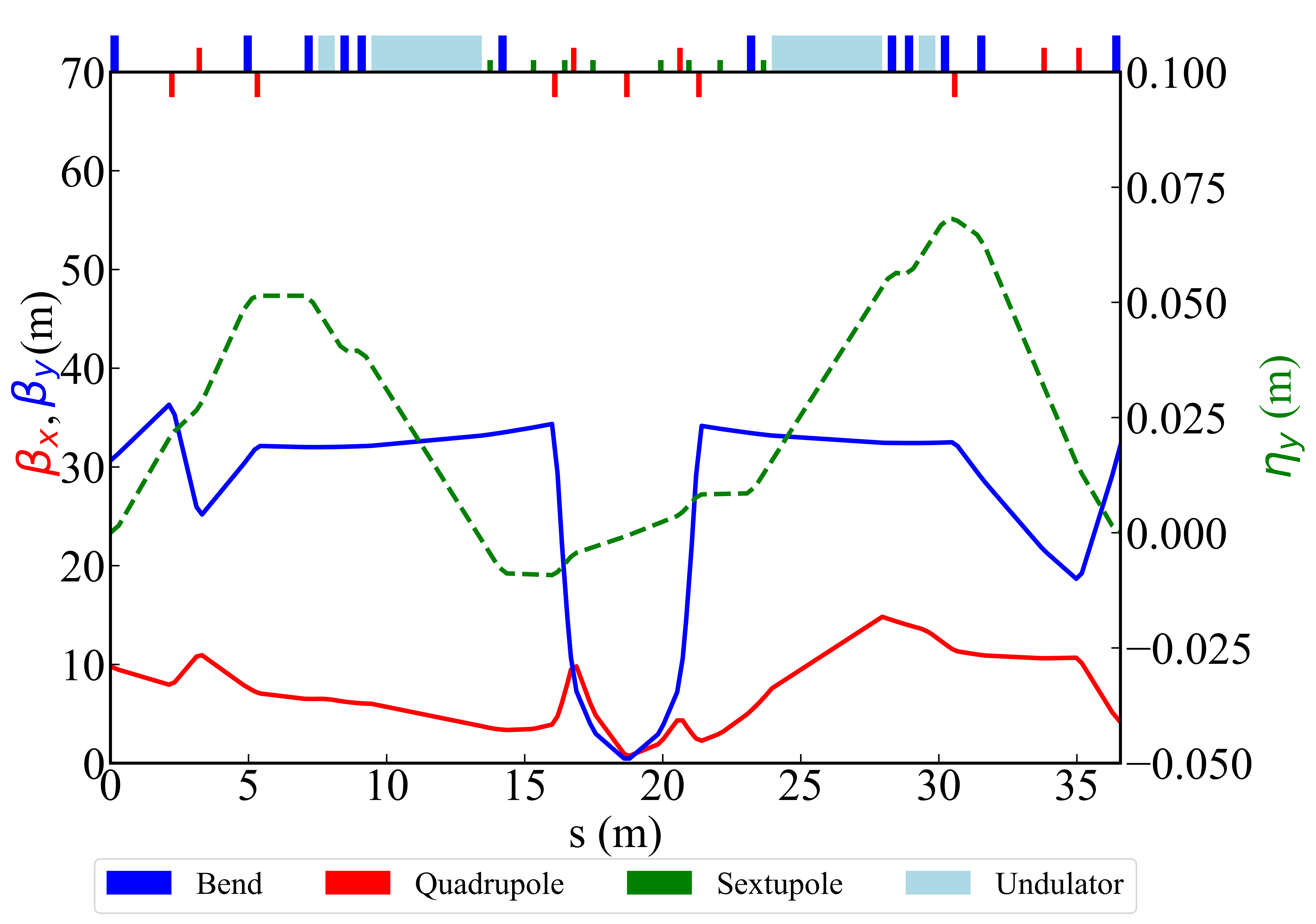}
\caption{Beam optics for bypass section.}
\label{fig:5}
\end{figure}
\begin{figure}[H]
\centering
\includegraphics[width=0.8\hsize]{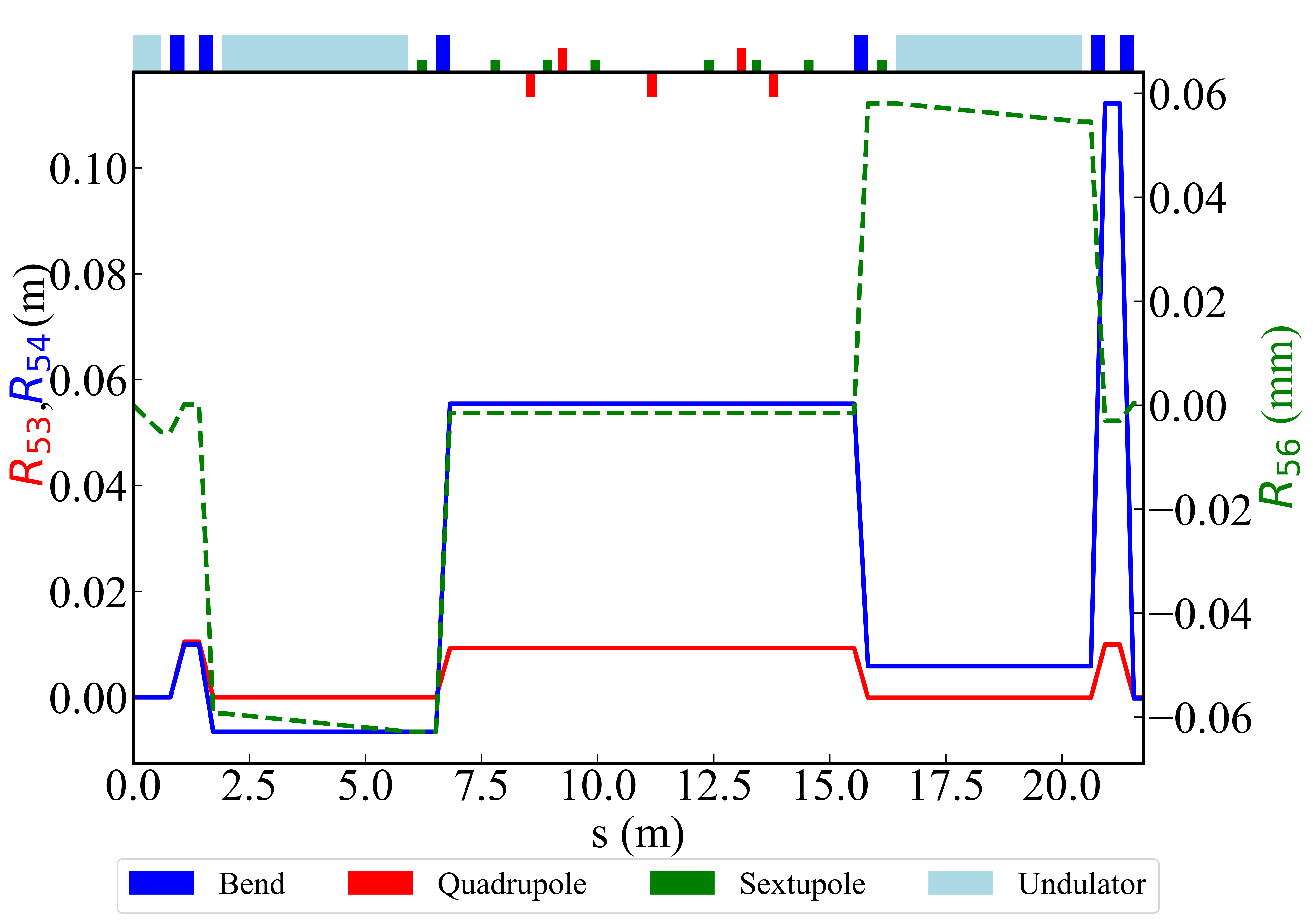}
\caption{The transfer matrix terms $R_{53}$, $R_{54}$ and $R_{56}$ between the entrance of the modulator and demodulator.}
\label{fig:6}
\end{figure}

Fig.~\ref{fig:5} show the optical functions of the bypass line. The first undulator serves as the energy modulation undulator in the ADM mechanism, while the second and third undulators are used as radiators. The last undulator is dedicated to energy demodulation. Between the two radiator sections, a double-bend structure is implemented, consisting of five quadrupoles and eight sextupoles. To preserve the longitudinal phase space of the electron beam between the entrances of the modulation and demodulation sections, the first-order transport matrix terms $R_{53},R_{54},R_{56}$ and the second-order terms $T_{511},T_{522},T_{521},T_{533},T_{544},T_{543},T_{566}$ should be optimized to approach zero. As shown in Fig.~\ref{fig:bypass}, the Dogleg 1 and Dogleg 2 sections provide identical values of $R_{54}$ and $R_{56}$, while the double-bend structure in the middle of the layout contributes $-2R_{54}$ and $-2R_{56}$. After the optimization, the energy spread introduced by modulation can be effectively canceled, thereby ensuring that the increase of emittance remains as small as possible. Fig.~\ref{fig:6} shows the values of $R_{53},R_{54},R_{56}$ between the centers of the modulation and demodulation sections. In addition, compared with Ref.~\cite{Jiang2022}, our bypass line design deliberately breaks the optical symmetry in order to provide higher flexibility in parameter optimization. Two matching sections are placed at the entrance and exit of the bypass line, each comprising two dipoles and three quadrupoles. Their purpose is to return the beam to the same vertical plane as the initial trajectory and to maintain achromatic conditions at both ends.
\subsection{Radiation performance}
Based on the design of the damping ring and the bypass line described above, this subsection focuses on the performance evaluation of coherent radiation. According to the damping-ring design described above, up to 200 bunches can be stored in the damping ring and can in principle be injected into the bypass line to produce radiation. The increase of the beam emittance and energy spread induced by the bypass line will be damped by the damping ring according to the following relation~\cite{DiMitri2021,Emma2001,Li2025}:
\begin{align}
\label{eq:41}
\varepsilon_{y}(t) &= \varepsilon_{y0} \exp\!\left(-\frac{2t}{\tau_{y}}\right) 
+ \left[ 1 - \exp\!\left(-\frac{2t}{\tau_{y}}\right) \right] \varepsilon_{ye}, \\
\label{eq:42}
\sigma_{\delta}^{2}(t) &= \sigma_{\delta 0}^{2} \exp\!\left(-\frac{2t}{\tau_{z}}\right) 
+ \left[ 1 - \exp\!\left(-\frac{2t}{\tau_{z}}\right) \right] \sigma_{\delta e}^{2},
\end{align}
where $\epsilon_{y0}$ and $\epsilon_{ye}$ are the vertical emittances after demodulation and at the equilibrium state, respectively; $\sigma_{\delta0}$ and $\sigma_{\delta e}$ are the energy spreads after demodulation and at the equilibrium state, respectively; $\tau_y$ and $\tau_z$ are the damping times in the vertical and longitudinal planes, respectively. According to the above equation, the electron beam experiences an increase in energy spread and emittance after passing through the bypass line, followed by damping in the damping ring. This cyclic process leads to a gradual growth of both emittance and energy spread until a new equilibrium state is reached. Since the radiation power based on the ADM mechanism is less sensitive to the beam energy spread than to the emittance, and as shown in Table~\ref{tab:1}, the transverse damping time is twice the longitudinal damping time, the primary concern is the beam emittance. According to the optimization condition of the ADM mechanism, when the initial beam emittance is 6 pm, the electron beam at the entrance of the radiator can form a well microbunching structure, thereby generating high-power coherent radiation. Therefore, the new equilibrium emittance is chosen to be 6 pm·rad. According to the bypass line design described in the previous section, the emittance of the electron beam increases from the initial value of 6 pm·rad to 6.109 pm·rad after passing through the bypass line, corresponding to a growth rate of 1.82$\%$. In contrast, the growth rate of the energy spread is only $5.38×10^{-5}$. The growth rate of the energy spread in the bypass line design is much smaller than that of the emittance. Therefore, the repetition rate of the coherent radiation is primarily determined by the vertical emittance damping. 

\begin{figure}[H]
\centering
\includegraphics[width=0.8\hsize]{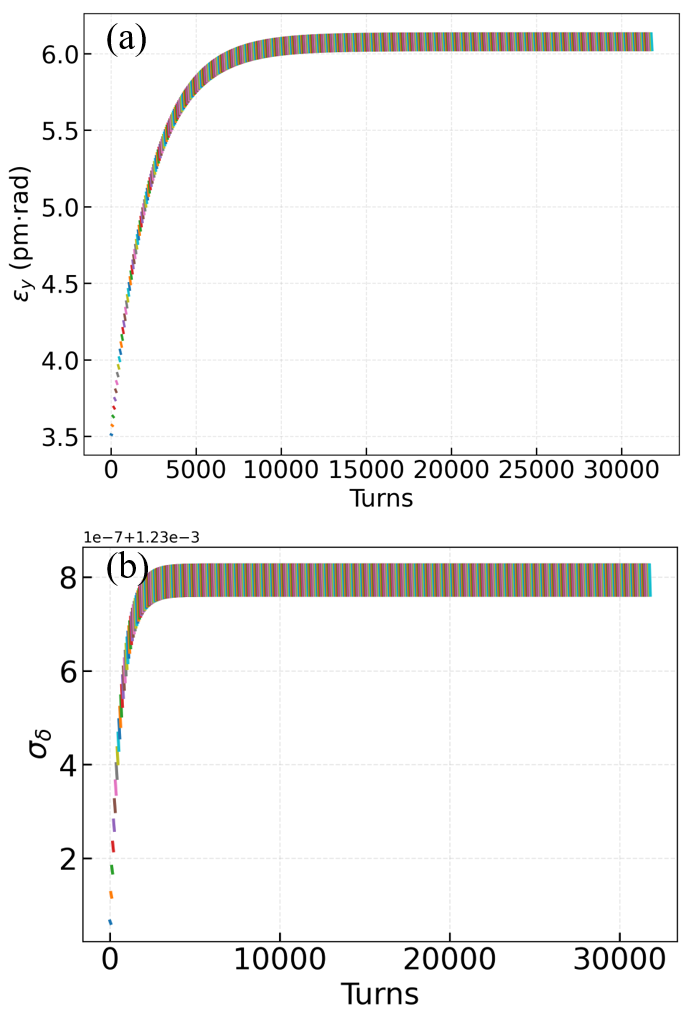}
\caption{(a) Vertical emittance evolution with the number of turns.(b) Energy spread evolution with the number of turns.}
\label{fig:7}
\end{figure}

\begin{figure}[H]
\centering
\includegraphics[width=0.8\hsize]{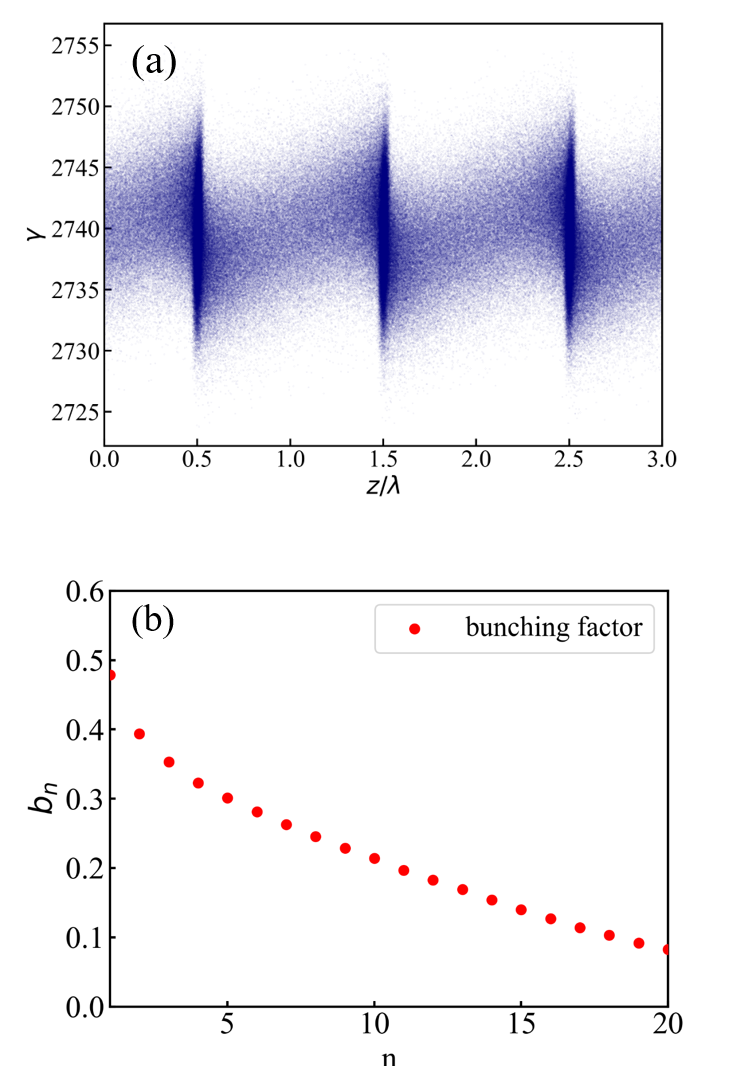}
\caption{(a) Longitudinal phase space distribution of the electron beam in the radiator and (b) the corresponding bunching factor.}
\label{fig:8}
\end{figure}

Substituting the above value together with the damping-ring parameters listed in Table~\ref{tab:1} into Eq.~\ref{eq:41} and ~\ref{eq:42}, the vertical emittance and energy spread evolution with the number of turns are obtained as shown in Fig.~\ref{fig:7}. We assume that after each modulation–demodulation in the bypass line, the electron beam undergoes 53 turns of damping in the storage ring. Fig.~\ref{fig:7} presents the results for the 600 modulation-demodulation. As shown, starting from the initial storage-ring vertical emittance of 3.45 pm·rad, the beam reaches a new equilibrium emittance of 6 pm·rad after multiple turns. Starting from the initial relative energy spread of $0.123\%$, the beam reaches a quasi-equilibrium energy spread of $0.123076\%$ after multiple turns. In the new equilibrium state, the energy spread is $0.123076\%$ and the emittance is 6 pm·rad. Based on the revolution frequency of our storage ring, the repetition rate of a single bunch is calculated to be 32.75 kHz. To ensure that the bunches can be extracted from the ring into the bypass line during the flat-top of the kicker pulse, we assume that 200 bunches are filled in the storage ring. The total radiation repetition rate reaches 6.55 MHz.
The next step is devoted to evaluating the single-pulse energy of coherent radiation. Following the optimized conditions of the ADM scheme, the corresponding parameters required to generate coherent radiation are obtained, with the main simulation parameters summarized in Table~\ref{tab:2}. 
\begin{table}[!htb]
\caption{bypass line parameters for simulation.}
\label{tab:2}
\begin{tabular*}{8cm} {@{\extracolsep{\fill} } lr}
\toprule
 Bending angle of  first bend (mrad) &  9.5  \\
 Length of first bend (m)    &   0.3   \\
 Laser wave length (nm)  &  266 \\ 
 Energy modulation amplitude ($\sigma_{E0}$) &  0.6 \\
 $R_{56}$ of dogleg ($\mu m$)      &   57   \\
 Dispersion of dogleg (mm)   &  6.5    \\
 Distance between two bends in dogleg (m)   &  0.314  \\
 Modulator length (m)   &  0.6  \\
 Modulator period length (m)   &  0.06  \\
 Radiator length (m)   &  4  \\
 Radiator period length (m)   &  0.025  \\
\bottomrule
\end{tabular*}
\end{table}
The electron beam parameters are primarily taken from Table~\ref{tab:1}. In the modulation section, The laser-electron beam interaction in the modulator induces an energy modulation amplitude of 0.6 times the intrinsic energy spread. Three-dimensional numerical simulations performed with Elegant show that the bunching factor at the 20th harmonic (13.3 nm) at the entrance of the radiator section reaches approximately 8.5$\%$, as illustrated in Fig.~\ref{fig:8},which is large enough to generate the coherent radiation. Fig.~\ref{fig:8}(a) presents the electron beam distribution at the radiator entrance, while Fig.~\ref{fig:8}(b) shows the corresponding bunching factor distribution.

\begin{figure}[H]
\includegraphics
  [width=0.9\hsize]{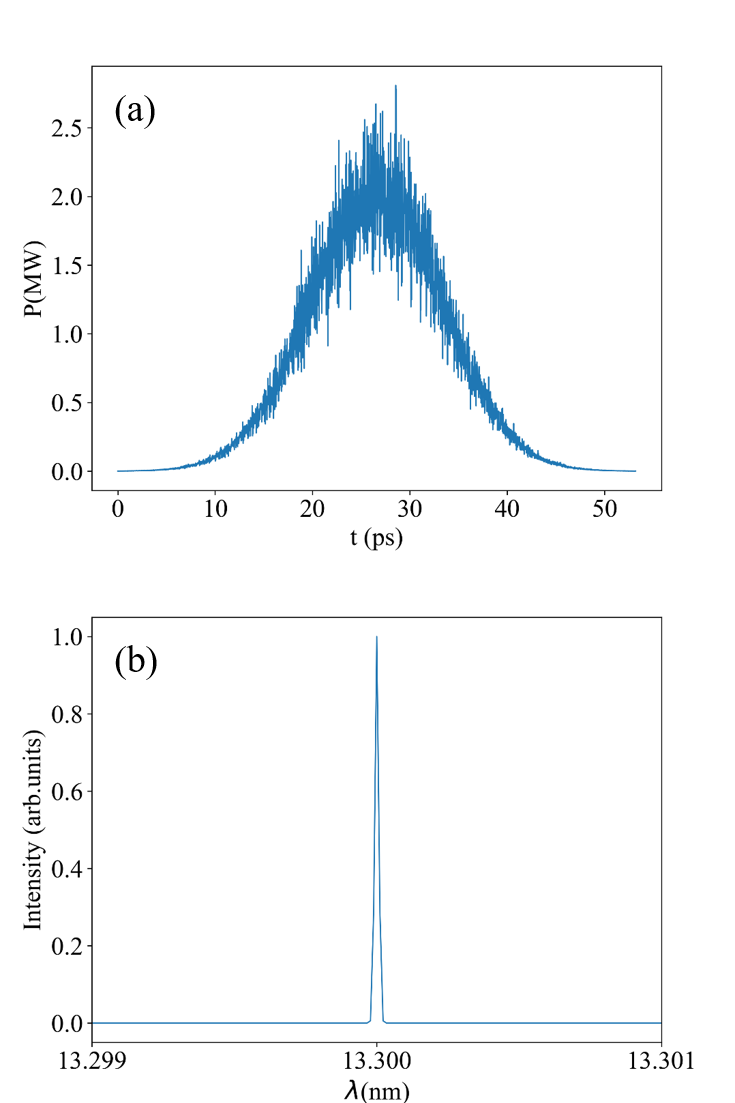}
\caption{Output radiation pulse and the corresponding single-shot spectrum.}
\label{fig:9}
\end{figure}
Finally, the coherent radiation process of the microbunched electron beam in the radiator was simulated using Genesis~\cite{Reiche1999}. Fig.~\ref{fig:9} shows the temporal profile of a single radiation pulse and the corresponding EUV spectrum. After passing through a 4-m-long undulator with a period length of 2.5 cm, a single-pulse energy of approximately 34.67 $\mu J$ can be generated. Combining this value with the radiation repetition rate of 6.55 MHz, the average EUV power is estimated to be about 227 W, exceeding the design target of 100 W.
\begin{figure}[H]
\centering
\includegraphics[width=0.8\hsize]{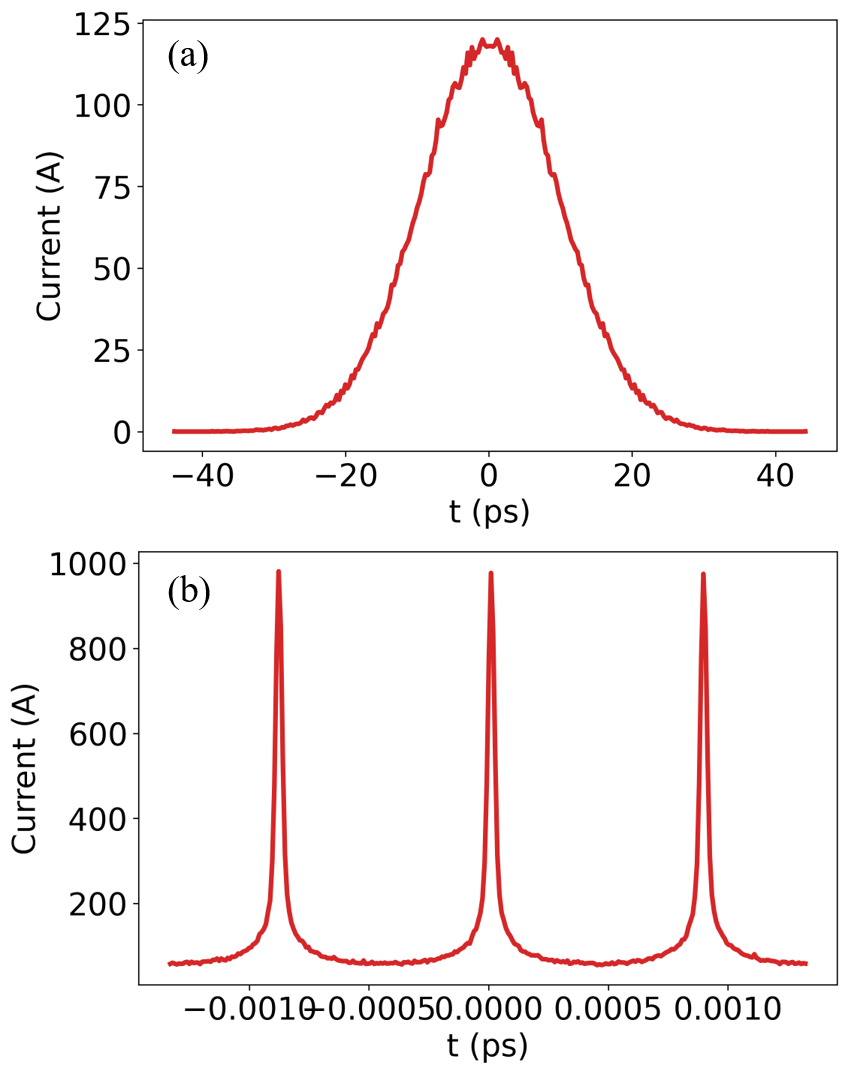}
\caption{(a) The current distribution of the entire bunch. (b) The current distribution of the microbunched beam formed through the ADM mechanism at the peak beam current.}
\label{fig:cur}
\end{figure}
Due to the modulation in the ADM within the bypass line, the electron beam locally forms microbunches, leading to a sharp increase in the local beam current. As shown in Fig.~\ref{fig:cur}(a), the current distribution of the entire bunch before modulation exhibits a peak current of about 120 A, while Fig.~\ref{fig:cur}(b) shows the current distribution after microbunching, where the peak current rises to nearly 1000 A. Therefore, it is of great importance to investigate the IBS effect in the bypass line. 

We use the IBSCATTER element in ELEGANT to simulate the impact of the IBS effect on the electron beam in the bypass line. The IBSCATTER element simulates IBS by tracking particles through the element, computing local scattering effects slice by slice, and accumulating the resulting changes to the beam’s emittance and energy spread over the process of the simulation. In the simulation, the electron beam is assumed to have an initial uniform current distribution of 120 A. We use $6.4×10^7$ macro-particles, longitudinally divided into 20,000 slices, with each slice being 13.3 nm long and containing 3200 particles. This setup ensures that, after microbunching, the electron beam exhibits the corresponding variations in current. Finally, when taking IBS into account, we tracked the electron beam through the entire bypass line and found that the emittance increased from 6 pm·rad to 6.1222 pm·rad, corresponding to a growth of 2.0367$\%$. Consistent with the previous analysis, the corresponding repetition rate is 5.98 MHz, resulting in an average power reduced to 200.77 W. It should be noted that only the coherent radiation from a single undulator is considered here.

\section{Conclusion}
In summary, we have proposed and systematically optimized a practical and robust lattice design for a fully storage-ring-based light source aimed at the generation of high-average-power EUV radiation. The ring adopts a four-superperiod structure. Each superperiod comprises four DBA cells and three high-field superconducting wigglers, providing strong radiation damping and a low natural emittance. In addition, the lattice exhibits excellent nonlinear dynamic performance, including a DA exceeding 17 mm and a minimum LMA greater than 3$\%$. Since this lattice extensively employs superconducting damping wigglers, the wiggler serves as a critical component of the lattice rather than merely an insertion device. To verify the reliability of the wiggler based on thin-dipole approximation, we developed an envelope-method code based on realistic field-map tracking. Then we verified its consistency with ELEGANT results, thereby ensuring the reliability of the design across the linear optics regime.

Furthermore, a new bypass line design with asymmetric optics was introduced. Based on the optimization conditions of the ADM mechanism, the single-pulse radiation energy was optimized by increasing the bunching factor. In addition, we minimize the emittance growth of the electron beam along the bypass line, thereby improving the demodulation performance. Meanwhile, the IBS effect on the microbunches was also taken into account in the bypass line. Finally, we demonstrated the feasibility of generating fully coherent EUV radiation with average output powers over hundred-watt. This combination of a strong-damping storage ring and a carefully tailored bypass line scheme highlights a promising pathway toward storage-ring-based coherent light sources. 

The influence of coherent radiation in the radiator on the electron beam is not been considered in this study. In this design, the repetition rate reaches 6.55 MHz, with a total of 200 bunches, meaning that repetition rate for single bunch is 32.75 kHz. This poses a significant challenge for the design of the kicker which is used to extract the bunches from the storage ring to the bypass. Both issues are currently under investigation.


\appendix
\section{thin-dipole Model}\label{app:dipole}
\begin{figure}[H]
\centering
\includegraphics[width=0.8\hsize]{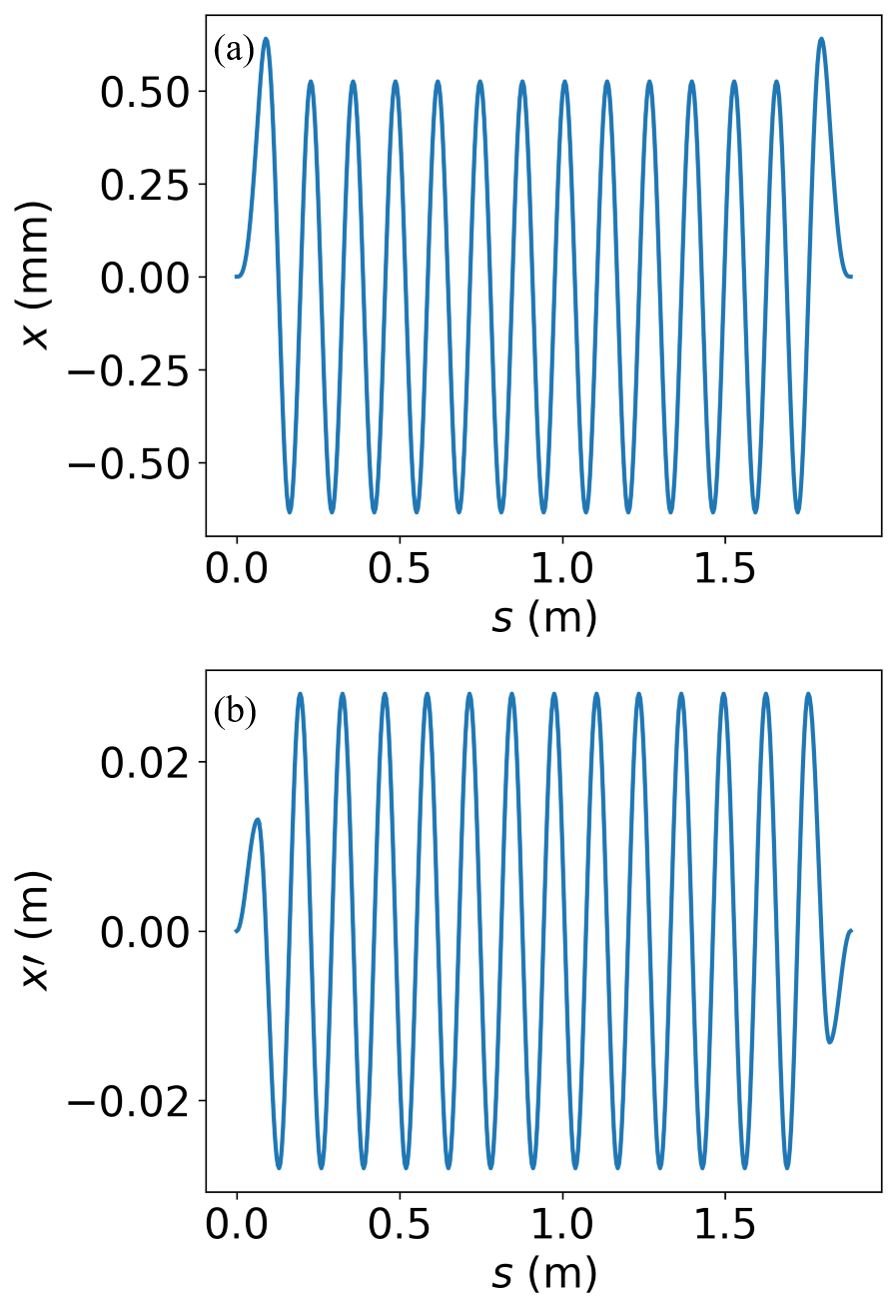}
\caption{Particles trajectories in the wiggler from implicit midpoint integrators.}
\label{fig:a11}
\end{figure}

\begin{figure}[H]
\centering
\includegraphics[width=0.8\hsize]{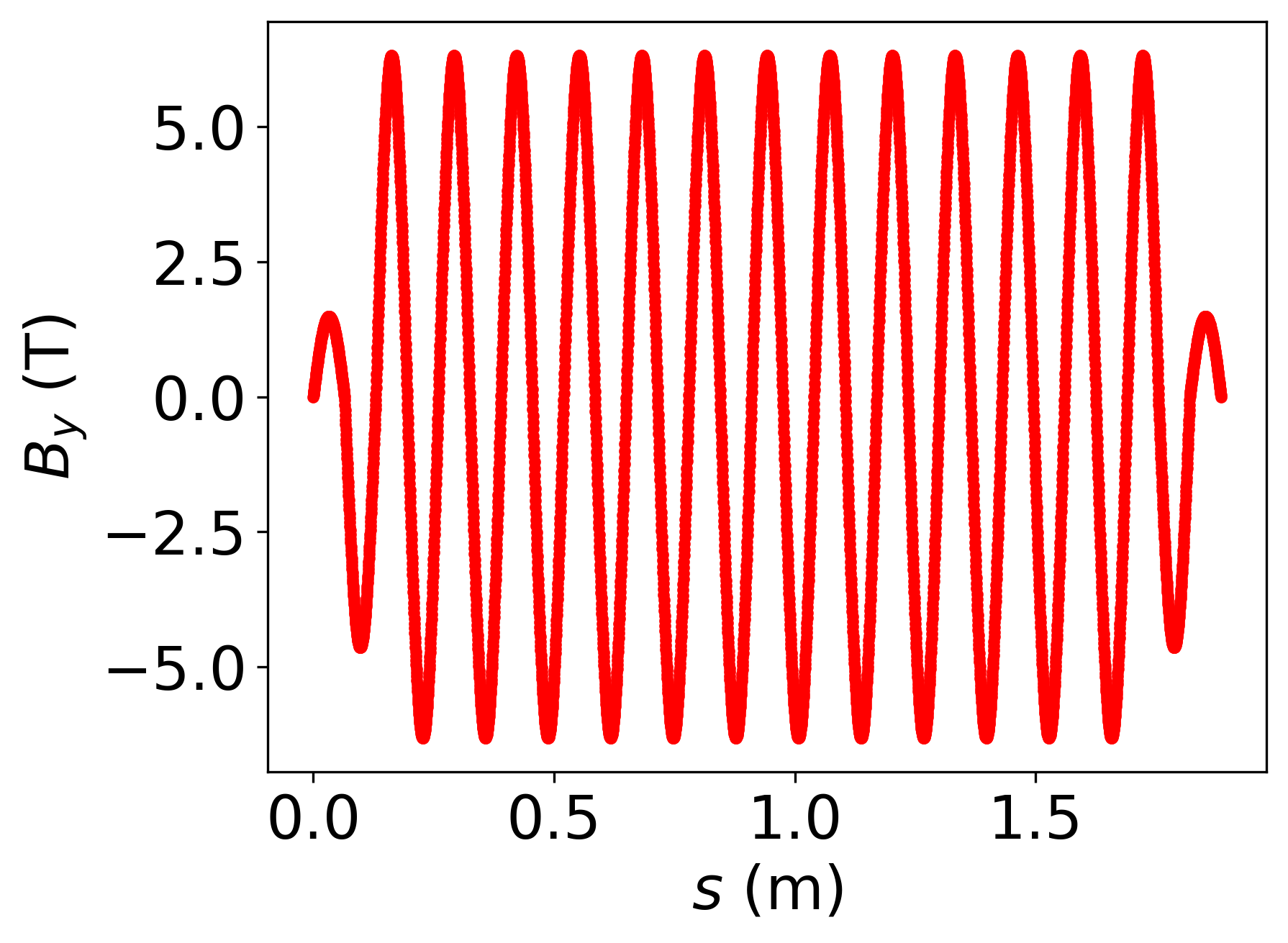}
\caption{Vertical magnetic field of the wiggler.}
\label{fig:by}
\end{figure}

To accurately evaluate the radiation damping and quantum excitation of the wigglers in the storage ring, the wigglers are modeled using number of thin dipoles. In this model, a wiggler is longitudinally divided into a number of thin dipole magnets. The Halbach expansions of the wiggler magnetic ﬁeld can be expressed as~\cite{Wolski2014,Wu2003,Li2019}:
\begin{equation}
    B_x = -\sum_{m,n}^{M,N} C_{mn} \frac{mk_x}{k_{y,mn}} \sin(mk_x x) \sinh(k_{y,mn} y) \sin(n k_z z),
    \label{eq:A3}
\end{equation}

\begin{equation}
    B_y = \sum_{m,n}^{M,N} C_{mn} \cos(mk_x x) \cosh(k_{y,mn} y) \sin(n k_z z),
    \label{eq:A4}
\end{equation}

\begin{equation}
    B_z = \sum_{m,n}^{M,N} C_{mn} \frac{n k_z}{k_{y,mn}} \cos(mk_x x) \sinh(k_{y,mn} y) \cos(n k_z z),
    \label{eq:A5}
\end{equation}
\begin{equation}
    k_{y,mn}^2 = m^2 k_x^2 + n^2 k_z^2.
    \label{eq:A6}
\end{equation}
We assume that the electron beam travels along the central axis of the wiggler ($y=0$). Consequently, the magnetic field components $B_x$ and $B_z$ vanish at $y=0$, and only the $B_y$ component contributes to the Hamiltonian. For simplicity, the $B_y$ can be expressed as:
\begin{equation}
    B_y=C_ysin(k_zz),
    \label{eq:A7}
\end{equation}
where $C_y$ is the peak magnetic field of the wiggler in one period. The peak magnetic fields at the ends of the wiggler are adjusted so that both the first and second integrals of the magnetic field are equal to 0. Then, this structure can be made achromatic while also allowing the electron beam to pass through transparently. After the optimization, the final design specifies a total wiggler length of 1.885 m, 14.5 periods, a period length of 130 mm, and a peak magnetic field of 6.3135 T. Fig.~\ref{fig:a11} shows the beam trajectory from Runge-Kutta method. 
Corresponding vertical magnetic field of the wiggler is shown in Fig.~\ref{fig:by}. Then, the magnetic field distribution of this wiggler can be represented using number of thin dipoles. The deflection angle of each dipole can be expressed as $L_bB_y/B\rho$, where $L_b$ is the length of the bend. The edge angle can be achieved through beam divergence trajectories.

\section{Envelope method based on wiggler in the storage ring}\label{app:envelope}
Here, we will introduce the envelope method based on wiggler in the storage ring. The matched lattice functions for the given storage ring configuration are obtained by first computing the closed orbit $\mathbf{x}_0(s)$ and the associated $6\times 6$ transfer matrices $M(s_j \leftarrow s_0)$ at a discrete set of locations $s_j$ around the ring. The closed orbit calculation ensures that all subsequent 
linear optics quantities are evaluated in a self-consistent reference frame 
where $\mathbf{x}_0(s)$ is periodic over one revolution.

The one-turn transfer map is defined as
\begin{equation}
\label{eq:B1}
M_C \equiv M(s_0+C \leftarrow s_0),
\end{equation}
where $C$ is the ring circumference. This matrix is diagonalized as
\begin{equation}
\begin{aligned}
M_C \, \mathbf{V}_k &= \lambda_k \, \mathbf{V}_k,\\
k&=I,II,III
\end{aligned}
\end{equation}
where the eigenvalues $\lambda_k$ occur in complex-conjugate pairs on the unit circle in the absence of damping, $\mathbf{V}_k$ is the eigenvector and $k$ is the eigenmode index. Each pair corresponds to one of the three independent oscillatory modes of the coupled $6$D system: horizontal betatron, vertical betatron and synchrotron motion.

The eigenvectors are normalized according to the symplectic form
\begin{equation}
\mathbf{V_k}^\dagger S \, \mathbf{V_k} = I,
\end{equation}
with the $6\times 6$ symplectic matrix
\begin{equation}
S = 
\begin{pmatrix}
0 & 1 & 0 & 0 & 0 & 0 \\
-1 & 0 & 0 & 0 & 0 & 0 \\
0 & 0 & 0 & 1 & 0 & 0 \\
0 & 0 & -1 & 0 & 0 & 0 \\
0 & 0 & 0 & 0 & 0 & 1 \\
0 & 0 & 0 & 0 & -1 & 0
\end{pmatrix}
\label{eq:skew_matrix}
\end{equation}
which defines the canonical coordinates $(x, p_x, y, p_y, z, \delta)$.
The mode identification is performed by projecting the normalized eigenvectors 
onto each canonical coordinate pair, thereby associating each eigenmode 
with the dominant subspace $(x, p_x)$, $(y, p_y)$, or $(z, \delta)$.

For each mode $k \in \{I, II, III\}$, we define the generalized Twiss matrix at the 
reference location $s_0$ as
\begin{equation}
\begin{aligned}
\beta_{ij}^k(0) &= \mathbf{V_k} \, T_k \, \mathbf{V_k}^\top,\\
i,j&=1,2,3,4,5,6,
\end{aligned}
\end{equation}
where 
\begin{subequations}\label{eq:matrices}
\begin{align}
T_I &= 
\begin{pmatrix}
0 & 1 & 0 & 0 & 0 & 0 \\
1 & 0 & 0 & 0 & 0 & 0 \\
0 & 0 & 0 & 0 & 0 & 0 \\
0 & 0 & 0 & 0 & 0 & 0 \\
0 & 0 & 0 & 0 & 0 & 0 \\
0 & 0 & 0 & 0 & 0 & 0
\end{pmatrix}, \label{eq:TI} \\
T_{II} &= 
\begin{pmatrix}
0 & 0 & 0 & 0 & 0 & 0 \\
0 & 0 & 0 & 0 & 0 & 0 \\
0 & 0 & 0 & 1 & 0 & 0 \\
0 & 0 & 1 & 0 & 0 & 0 \\
0 & 0 & 0 & 0 & 0 & 0 \\
0 & 0 & 0 & 0 & 0 & 0
\end{pmatrix}, \label{eq:TII} \\
T_{III} &= 
\begin{pmatrix}
0 & 0 & 0 & 0 & 0 & 0 \\
0 & 0 & 0 & 0 & 0 & 0 \\
0 & 0 & 0 & 0 & 0 & 0 \\
0 & 0 & 0 & 0 & 0 & 0 \\
0 & 0 & 0 & 0 & 0 & 1 \\
0 & 0 & 0 & 0 & 1 & 0
\end{pmatrix}. \label{eq:TIII}
\end{align}
\end{subequations}
$T_k$ is the $6\times 6$ projector onto the canonical subspace of mode $k$. The generalized Twiss matrices are propagated along the storage ring using
\begin{equation}
\label{eq:betaa}
\beta_{ij}^k(s_n) = M(s_n \leftarrow s_0) \, \beta_{ij}^k(0) \, M(s_n \leftarrow s_0)^\top.
\end{equation}
Simultaneously, we construct a real canonical transformation matrix
\begin{equation}
N = \sqrt{2} \big[ \mathscr{R}(v_1), \mathscr{I}(v_1), \mathscr{R}(v_3), \mathscr{I}(v_3), \mathscr{R}(v_5), \mathscr{I}(v_5)],
\end{equation}
where $\mathscr{R}(v_1)$, $\mathscr{R}(v_3)$ and $\mathscr{R}(v_5)$ denote the real parts of the first, third, and fifth columns of the normalized eigenvectors, while $\mathscr{I}(v_1)$, $\mathscr{I}(v_3)$ and $\mathscr{I}(v_5)$ represent their corresponding imaginary parts. For each location $s_n$, the local phase advances $\mu_k(s_n)$ are extracted from the block-diagonal form of the similarity transform:
\begin{equation}
 R(s_n) = N(s_n)^{-1} \, M(s_n \gets s_{n-1}) \, N(s_{n-1}),
\end{equation}
where each 2×2 block in R is a pure rotation of the form:
\begin{equation}
 R_k = \begin{pmatrix} 
\cos \mu_k & \sin \mu_k \\ 
-\sin \mu_k & \cos \mu_k 
\end{pmatrix}. 
\end{equation}
The full ring tunes are then given by
\begin{equation}
\label{eq:15}
\nu_k(s_n) = \frac{1}{2\pi} \, \mu_k(s_n), \quad k=I,II,III,
\end{equation}
with the one-turn tunes $\nu_k(C)$ obtained by evaluating at $s=C$.

In the uncoupled limit, $\beta_k$ reduces to the familiar Courant–Snyder form. The familiar Twiss parameters in the horizontal plane can be expressed as:
\begin{subequations}\label{eq:beta1}
\begin{align}
\beta_x&=\beta^I_{11},\\
\alpha_x&=-\beta^I_{12},\\
\gamma_x&=\beta^I_{22}.
\end{align}
\end{subequations}
In the horizontal plane, they can be expressed as:
\begin{subequations}\label{eq:beta2}
\begin{align}
\beta_y&=\beta^{II}_{33},\\
\alpha_y&=-\beta^{II}_{34},\\
\gamma_y&=\beta^{II}_{44}.
\end{align}
\end{subequations}
Finally, the dispersion in the horizontal and vertical plane can be expressed as:
\begin{subequations}\label{eq:eta}
\begin{align}
\eta_x&=\frac{\beta^{III}_{16}}{\beta^{III}_{66}},\\
\eta_y&=\frac{\beta^{III}_{36}}{\beta^{III}_{66}}.
\end{align}
\end{subequations}

In the next step, we will employ the envelope method in conjunction with the wiggler field map to compute the equilibrium emittance of the storage-ring light source. First, we will introduce the covariance matrix to describe the beam distribution:
\begin{equation}
\Sigma = \begin{pmatrix}
\langle x^2 \rangle      & \langle x p_x \rangle & \langle x y \rangle   & \langle x p_y \rangle & \langle x z \rangle   & \langle x \delta \rangle \\
                         & \langle p_x^2 \rangle & \langle p_x y \rangle & \langle p_x p_y \rangle & \langle p_x z \rangle & \langle p_x \delta \rangle \\
                         &                       & \langle y^2 \rangle   & \langle y p_y \rangle & \langle y z \rangle   & \langle y \delta \rangle \\
                         &                       &                       & \langle p_y^2 \rangle & \langle p_y z \rangle & \langle p_y \delta \rangle \\
                         &                       &                       &                       & \langle z^2 \rangle   & \langle z \delta \rangle \\
                         &                       &                       &                       &                       & \langle \delta^2 \rangle
\end{pmatrix}.
\end{equation}
This can be compactly written as
\begin{equation}
\Sigma_{ij}=\langle X_iX_j \rangle, \vec{X}^{\mathrm{T}} = \begin{pmatrix}
x & p_x & y & p_y & z & \delta
\end{pmatrix},
\end{equation}
where $\Sigma_{ij}$ is the $(i,j)$ component of the "Sigma matrix", and the $X_{i,j} (i,j=1...6)$ are the dynamical variables. The brackets $\langle\cdot\rangle$ indicate an average overall particles in the bunch. Neglecting radiation and other effects, and considering only the Lorentz force from external electromagnetic fields, the transport remains symplectic. According to the one-turn transfer map in Eq.~\ref{eq:B1}, the equilibrium beam distribution can be described as:
\begin{equation}
\Sigma_{ij}(s_0)=\Sigma_{ij}(s_0+C)=\langle M(X_i)M(X_j) \rangle.
\end{equation}
Therefore, the matched distribution is one that satisfies:
\begin{equation}
M\cdot\Sigma\cdot M^T=\Sigma.
\end{equation}
However, in a storage ring, the presence of radiation leads to the damping of the emittance toward an equilibrium value. As a consequence, the previously defined transfer matrix $M$ is no longer symplectic. Moreover, in addition to the first-order terms, the transfer matrix $M$ also contains zero-order contributions arising from quantum excitation. Therefore, a matched distribution should then be written as:
\begin{equation}
\label{eq:B19}
\Sigma=M\cdot\Sigma\cdot M^T+D.
\end{equation}
The first step of the envelope method for calculating the equilibrium emittances in a storage ring is to find the first-order terms $M$ and zeroth-order terms $D$. For the transfer matrix $M$, in addition to the conventional elements associated with magnetic optics, the radiation damping terms must also be included. The matrix $D$, on the other hand, primarily accounts for the quantum excitation terms. In our storage-ring light source, the elements that introduce radiation effects are mainly the RF cavities, dipole magnets, and superconducting wigglers. In particular, the quantum excitation terms for the dipole magnets and wigglers can be expressed as:
\begin{align}
D_{66}(\text{bend})
      &= \frac{2 C_L \gamma^5}{c |\rho|^3}, \\[6pt]
D_{66}(\text{wiggler})
      &= \frac{2 C_L \gamma^5}{c |B\rho/B_y|^3},
\end{align}
where c is the speed of light in
free space,$B_y$ is the vertical magnetic field of the wiggler, $B\rho$ is the magnet rigidity. Their damping terms can be expressed as:
\begin{align}
M_{66} (\text{bend}) &= - \frac{C_{\gamma} E_{0}^{3}}{\pi} \frac{1}{\rho^{2}}, 
\\
M_{61} (\text{bend}) &= - \frac{C_{\gamma} E_{0}^{3}}{2\pi} \frac{1}{\rho^{3}},\\
M_{66} (\text{wiggler}) &= - \frac{C_{\gamma} E_{0}^{3}}{\pi} \frac{1}{(B\rho/B_y)^{2}}, 
\\
M_{61} (\text{wiggler}) &= - \frac{C_{\gamma} E_{0}^{3}}{2\pi} \frac{1}{(B\rho/B_y)^{3}},
\end{align}
where $C_\gamma = \frac{4\pi}{3}\frac{r_e}{(mc^2)^3} = 8.85\times 10^{-5} \frac{m}{GeV^2}$. For an RF cavity, we have all the other damping matrix terms of M zero except that
\begin{equation}
M_{22} = M_{44} = - \frac{e V_{RF} \sin \phi_{RF}}{E_{0}} \,
\end{equation}
 where $e$ is the elementary charge, $V_{RF}$ and $\phi_{RF}$ are the RF voltage and phase, respectively. Using the matching condition Eq.~\ref{eq:B19}, we will determine the Sigma matrix. Given the Sigma matrix at a location $s_0$, the Sigma matrix at a location $s_1=s_0+ds$ can be expressed as:
\begin{equation}
\Sigma(s_{1}) = M(s_{1}\leftarrow s_{0}) \cdot \Sigma(s_{0}) \cdot M^{\top}(s_{1}\leftarrow s_{0}) + D(s_{1} \leftarrow s_{0})
\end{equation}
The Sigma matrix at $s_2$ is given by:
\begin{widetext}
\begin{align}
\Sigma(s_{2}) &= M\bigl(s_{2}\leftarrow s_{1}\bigr) \cdot \Sigma(s_{1}) \cdot M^{\top}\bigl(s_{2}\leftarrow s_{1}\bigr) + D\bigl(s_{2}\leftarrow s_{1}\bigr) \notag \\
              &= M\bigl(s_{2}\leftarrow s_{0}\bigr) \cdot \Sigma(s_{0}) \cdot M^{\top}\bigl(s_{2}\leftarrow s_{0}\bigr) \notag \\
              &\quad + M\bigl(s_{2}\leftarrow s_{1}\bigr) \cdot D\bigl(s_{1}\leftarrow s_{0}\bigr) \cdot M^{\top}\bigl(s_{2}\leftarrow s_{1}\bigr) + D\bigl(s_{2}\leftarrow s_{1}\bigr).
\end{align}
\end{widetext}
Then:
\begin{widetext}
\begin{align}
M(s_2\leftarrow s_0) &=
M(s_2\leftarrow s_1)
M(s_1\leftarrow s_0) \\
D(s_2\leftarrow s_0) &=M(s_2\leftarrow s_1)
D(s_1\leftarrow s_0)M^T(s_2\leftarrow s_1)+D(s_2\leftarrow s_1)
\end{align}
\end{widetext}
Then, we can achieve:
\begin{widetext}
\begin{align}
M(s_n\leftarrow s_0) &=
M(s_n\leftarrow s_{n-1})
M(s_{n-1}\leftarrow s_{n-2})...M(s_1\leftarrow s_0) \\
D(s_n\leftarrow s_0) &=\sum_{r=1}^{n}M(s_2\leftarrow s_r) D(s_r \leftarrow s_{r-1})M^T(s_n\leftarrow s_r)
\end{align}
\end{widetext}

Finally, we can find the equilibrium emittances by solving Eq.~\ref{eq:B19}. Making use of the eigenvectors $U$ of $M$, we can construct the diagonal matrix $\Lambda$ from the eigenvalues of $M$:
\begin{align}
M\cdot U =\Lambda\cdot U.
\end{align}
We can define $\tilde{\Sigma}$ and $\tilde{D}$ as:
\begin{align}
\Sigma&=U\tilde{\Sigma}\cdot U^T,\\
D&=U\tilde{D}\cdot U^T.
\end{align}
The solution for the Sigma matrix can be written as:
\begin{align}\label{eq:sigma}
\tilde{\Sigma}_{ij} = \frac{\tilde{D}_{ij}}{1 - \Lambda_i \Lambda_j}.
\end{align}
The matched (equilibrium) beam distribution $\Sigma$ can be obtained. The corresponding emittances are then determined from the three eigenvalues of $\Sigma S$.

\end{document}